\documentclass[conference]{IEEEtran}

\IEEEoverridecommandlockouts

\usepackage{amsmath,amssymb,amsfonts}
\usepackage{mathtools}
\usepackage{mathabx}
\usepackage{amsthm}
\usepackage{nicefrac
}

\usepackage{graphicx}
\usepackage{textcomp}
\usepackage[table]{xcolor}
\def\BibTeX{{\rm B\kern-.05em{\sc i\kern-.025em b}\kern-.08em
    T\kern-.1667em\lower.7ex\hbox{E}\kern-.125emX}}

\usepackage[hidelinks]{hyperref}
\usepackage{subfig}
\usepackage{tikz}
\usepackage[nolist,nohyperlinks]{acronym}
\usepackage{siunitx}
\usepackage{cite}
\usepackage{microtype}
\usepackage{nicefrac}
\renewcommand{\baselinestretch}{0.9} 
\usepackage{microtype}
\definecolor{mylightgreen}{RGB}{245,248,242}
\definecolor{mydarkgreen}{RGB}{70,170,70}

\DeclareSIUnit\dyne{dyn}
\DeclareSIUnit\poise{P}
\DeclareSIUnit{\molar}{M}
\DeclareSIUnit{\particle}{particles}

\makeatletter
\newcommand\footnoteref[1]{\protected@xdef\@thefnmark{\ref{#1}}\@footnotemark}
\makeatother

\newcommand{\defeq}{\vcentcolon=}

\acrodef{0-D}[0-D]{zero-dimensional}
\acrodef{1-D}[1-D]{one-dimensional}
\acrodef{2-D}[2-D]{two-dimensional}
\acrodef{3-D}[3-D]{three-dimensional}
\acrodef{CVS}[CVS]{cardiovascular system}
\acrodef{VN}[VN]{vessel network}
\acrodef{LS}[LS]{lymphatic system}
\acrodef{MC}[MC]{molecular communications}
\acrodef{ODE}[ODE]{ordinary differential equation}
\acrodef{CGS}[CGS]{centimetre-gram-second}
\acrodef{HEV}[HEV]{high endothelial venule}
\acrodef{TNFalpha}[TNF-$\alpha$]{Tumor Necrosis Factor alpha}
\acrodef{IL1beta}[IL-1$\beta$]{Interleukin-1 beta}
\acrodef{IL6}[IL-6]{Interleukin 6}
\acrodef{PBS}[PBS]{particle-based simulation}
\acrodef{HIV}[HIV]{Human Immunodeficiency Virus}
\acrodef{SNR}[SNR]{signal-to-noise ratio}
\acrodef{IID}[IID]{independent and identically distributed}
\acrodef{Rx}[Rx]{receiver}
\acrodef{RV}[RV]{random variable}
\acrodef{ICG}[ICG]{indocyanine green}
\acrodef{ROI}[ROI]{region of interest}
\acrodefplural{PBS}[PBSs]{particle-based simulations}
\acrodefplural{HEV}[HEVs]{high endothelial venules}
\acrodefplural{VN}[VNs]{vessel networks}
\acrodefplural{ROI}[ROIs]{regions of interest}
\acrodefplural{Rx}[Rxs]{receivers}
\acrodefplural{RV}[RVs]{random variables}

\begin{document}
\bstctlcite{IEEEexample:BSTcontrol}  % supress em-dash for repeating authors

\title{%
Simulation and Analysis of Solute Transport in Multi-Lymphangion Lymphatic Vessels
\thanks{%
This work was funded by the German Federal Ministry of Research, Technology and Space through Project Internet of Bio-Nano-Things -- grant number 16KIS1987, by the German Research Foundation under GRK 2950 -- ProjectID 509922606 and under grant number SCHA 2350/2-1, by the European Union’s Horizon Europe -- HORIZON-EIC-2024-PATHFINDEROPEN-01 under grant agreement Project N. 101185661, and by the Horizon Europe Marie Skodowska Curie Actions (MSCA)-UNITE under Project 101129618.}
}

\author{\IEEEauthorblockN{\scalebox{.99}{%
  Timo Jakumeit$^\text{1}$, Thiha Aung$^\text{2}$, Robert Schober$^\text{1}$, and Maximilian Sch\"afer$^\text{1}$%
}}\\[-0.4cm]
\IEEEauthorblockA{\small $^\text{1}$Friedrich-Alexander-Universität Erlangen-Nürnberg (FAU), Erlangen, Germany\\ $^\text{2}$Deggendorf Institute of Technology, Deggendorf, Germany}\vspace*{-7mm}
}

\maketitle

\begin{abstract}
The \ac{LS}, a body-wide network of vessels and lymphoid organs governing fluid homeostasis and immune surveillance, has so far not been investigated as a domain for diagnostic and therapeutic \ac{MC} applications, despite several properties that make it a promising, complementary alternative to the cardiovascular system.
These favorable properties include slower flow, simpler and less dense molecular fluid composition, and direct anatomical access to lymph nodes.
Realizing this potential, however, requires a quantitative understanding of how solutes propagate through the \ac{LS}, a problem that, unlike lymph flow itself, remains largely unaddressed in the literature.
As a first step towards narrowing this gap, we develop a \ac{PBS} framework of solute transport through a three-dimensional chain of valve-separated, concatenated vessel segments, called \textit{lymphangions}. 
We simulate the spatiotemporal evolution of solute concentration and qualitatively validate the resulting transport dynamics against existing \textit{in vivo} measurements of fluorescent tracer propagation in multi-lymphangion lymphatic vessels. 
Our simulations show that the valve-gated nature of solute transport in lymphatic vessels leads to bursty solute concentrations over time, a characteristic that can also be observed \textit{in vivo}.
Additionally, we find that, within one pumping period, peak timing is dictated by the valves' synchronizing effect rather than the particle release time, while diffusivity and receiver placement determine peak sharpness.
Overall, the proposed \ac{PBS} framework provides a first quantitative basis for solute transport modeling in the \ac{LS} and several concrete application scenarios for \ac{MC} in this underexplored domain.
Supplementary video material illustrating the \ac{PBS} is publicly available on Zenodo [DOI: \url{10.5281/zenodo.21888066}].
\end{abstract}

\acresetall

\iffalse
\fi

\section{Introduction}\label{sec:Introduction}
Every day, the human body filters roughly eight to twelve liters of fluid out of the bloodstream and into the surrounding tissue; fluid that must be returned to circulation or the body would swell fatally within hours~\cite{Breslin2019}.
This task falls to the \ac{LS}, a body-wide network of vessels and lymphoid organs that performs fluid homeostasis, immune surveillance, and waste clearance in parallel to the \ac{CVS}.
Despite its scale and centrality, the \ac{LS} has historically received less scientific attention than the \ac{CVS}~\cite{Bertram2010}, in part because \textit{lymph} flow is difficult to observe and measure directly~\cite{Sharma2007}, and in part because the lymphatic vasculature lacks a central pump analogous to the heart, instead relying on a chain of self-contractile segments called \textit{lymphangions} to actively propel lymph forward~\cite{Breslin2019}. This makes the \ac{LS} considerably harder to characterize than the \ac{CVS}.

A similar imbalance has emerged in the field of \ac{MC}: numerous health-monitoring and therapeutic applications have been envisioned within the \ac{CVS}, owing to its pervasiveness in the body and its efficient transport of signaling molecules, while the \ac{LS}, despite brief mentions in~\cite{Barros2021, Felicetti2016}, has, to the best of our knowledge, not yet been considered in its own right as an application domain of \ac{MC}.

However, several properties of the \ac{LS} suggest that it may be a more suitable, or at least complementary, domain for \ac{MC}-based healthcare applications~\cite{Breslin2019}:
(i)~The \ac{LS} exhibits slower fluid flow, relaxing timing and sampling-rate requirements for sensors.
(ii)~Lymph carries a far less dense and complex mixture of proteins and cells than blood, reducing background noise for target molecule detection.
(iii)~Lymph composition directly reflects the immune and inflammatory state of the tissue it drains, making the \ac{LS} an information-rich channel for health monitoring.
(iv)~The \ac{LS} offers a direct anatomical route to \textit{lymph nodes}, which are key therapeutic targets that the \ac{CVS} reaches only indirectly and inefficiently.

\begin{figure*}
    \centering
    \includegraphics[width=\textwidth]{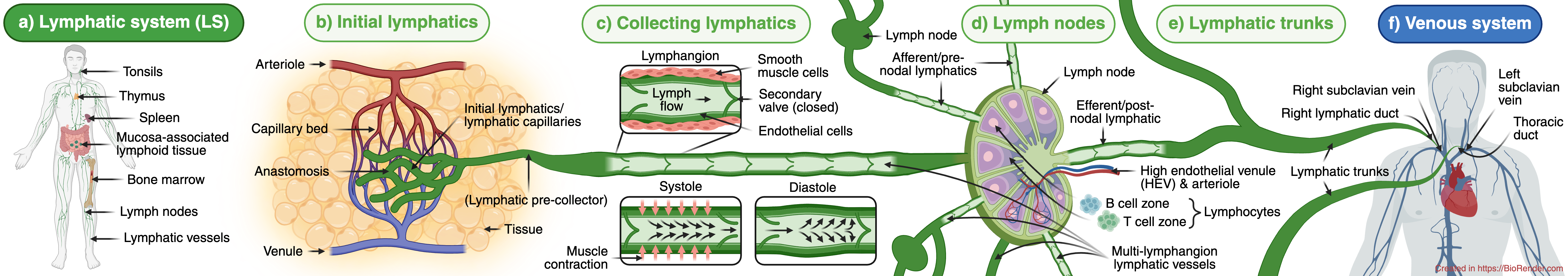}\vspace*{1mm}
    \caption{\textbf{Overview of the \ac{LS} anatomy.} \textbf{a)}~Main components of the \ac{LS}. \textbf{b)}~Initial lymphatics drain interstitial fluid from peripheral tissue. \textbf{c)}~Collecting lymphatics actively propel lymph forward through self-contracting, valve-separated lymphangions. \textbf{d)}~Afferent collecting lymphatics deliver lymph to lymph nodes, where its contents encounter lymphocytes of the immune system. \textbf{e)}~Lymphatic trunks carry lymph toward the venous system, \textbf{f)} returning it to the \ac{CVS} via lymphatic ducts.}
    \label{fig:LS_overview}
\end{figure*}

Realizing these advantages for \ac{MC}, however, first requires a quantitative understanding of the \ac{LS} via suitable models.
In the field of biomechanics, several numerical and analytical models have been proposed for lymph flow in lymphangions, lymph nodes, and lymphatic \acp{VN}, and for the network topology of the \ac{LS}. 
The authors of~\cite{Bertram2010,Bertram2013} present an analytical model for lymph flow in a chain of lymphangions and analyze the influence of system parameters on pumping effectiveness. 
Lymph flow in lymph nodes has been modeled using finite-element and -volume solvers, e.g., in~\cite{Cooper2015, Giantesio2022}, with~\cite{ Giantesio2022} additionally providing an analytical model for a simplified node geometry. 
At the \ac{VN} scale, the authors of~\cite{ Savinkov2020} propose a graph representation of all major lymphatic vessels and nodes in the human body and analyze topological features, while the authors of~\cite{Jamalian2016} derive a lymph flow model for lymphatic \acp{VN} and study the effect of system parameters on pumping behavior.
Crucially, as remarked in~\cite{JayathungageDon2023}, models or simulations for the \textit{transport of solutes} within chains of lymphangions, lymph nodes or other parts of the \ac{LS} are scarce, pointing towards a critical research gap.
To the best of the authors' knowledge, only three works in this direction exist to date: 
First, in \cite{Li2024}, a lattice Boltzmann simulation for the transport of a single leukocyte in a lymphangion chain is proposed. 
Second, \cite{Han2023} proposes a poroelastic model for the delivery of subcutaneously injected solutes through soft skin tissue to the initial lymphatics.
Third, in \cite{Medina2022}, a \ac{2-D} fluid-structure interaction model is proposed for the transport of lymphocytes through a single valved lymphangion.

To narrow this gap, in this work, we develop a \ac{3-D} \ac{PBS} framework to simulate and analyze the transport of particles suspended in lymph as they are advected through a chain of concatenated lymphangions.
The underlying lymph flow is prescribed by the \ac{0-D} model established in \cite{Bertram2010}. 
To characterize the spatiotemporal evolution of transported solutes, we place transparent \acp{Rx} along the lymphatic vessel and analyze the observed concentration profiles over time.
The simulated transport dynamics are then qualitatively compared with \textit{in vivo} measurements of tracer molecule distributions obtained from multi-lymphangion lymphatic vessels in the abdomen of a swine model~\cite{Sharma2007}.
 
The contributions of this work are as follows:
\begin{enumerate}
    \item We propose the \ac{LS} as a domain for health-related \ac{MC} applications and, for the first time in the \ac{MC} literature, outline several concrete application scenarios.
    \item Building on the lymph flow model for chains of lymphangions in~\cite{Bertram2010, Bertram2013}, we develop a \ac{PBS} framework for solute transport in multi-lymphangion lymphatic vessels.
    \item We study the impact of various system parameters on particle transport and qualitatively validate the \ac{PBS} results against \textit{in vivo} measurements of \ac{ICG} propagation in lymphatic vessels~\cite{Sharma2007}.
\end{enumerate}

The remainder of this paper is organized as follows:
Section~\ref{sec:Overview_Lymphatic_System} provides an overview of the \ac{LS}.
Envisioned applications of \ac{MC} in the \ac{LS} are proposed in Section~\ref{sec:Applications}, followed by an introduction to the lymph flow model in~\cite{Bertram2010} in Section~\ref{sec:LymphFlowModel}.
A \ac{PBS} framework for particle transport in lymphangion chains is proposed in Section~\ref{sec:MoleculeTransportSimulation} and parameter sweeps as well as a comparison of numerical results and \textit{in vivo} measurements from~\cite{Sharma2007} are presented in Section~\ref{sec:NumericalResults}.
Section~\ref{sec:Conclusion} concludes the paper.
Supplementary video material illustrating the proposed \ac{PBS} is publicly available on Zenodo, cf.~\cite{jakumeit2026zenodo}.

\section{Overview of the Lymphatic System}\label{sec:Overview_Lymphatic_System}

Of all major organ systems, the \ac{LS} is most intimately tied to the \ac{CVS} and immune system, yet it has historically received far less research attention than either~\cite{Bertram2013, Breslin2019}. 
We therefore provide a brief overview of its function and structure here; for a comprehensive treatment, we refer to~\cite{Breslin2014, Breslin2019}.
A schematic of the \ac{LS} is shown in Fig.~\ref{fig:LS_overview}. 

The \ac{LS} serves several interrelated functions: 
maintaining the homeostasis of extracellular fluid for optimal tissue function, supporting the immune system by providing pathways for pathogen sensing and immune signaling, and clearing metabolic waste products and cellular debris resulting from apoptosis and normal metabolism~\cite{Breslin2019}. 
To fulfill these roles, the \ac{LS} roughly mirrors the spatial organization of the \ac{CVS}, extending as a \ac{VN} through most regions of the human body, see Fig.~\ref{fig:LS_overview}a).

Proceeding from smallest to largest scale, the system begins with \textit{initial lymphatics}, i.e., blind-ended, thin-walled vessels consisting of a single endothelial cell layer, that form anastomotic \acp{VN} interfacing with cardiovascular capillary beds, see Fig.~\ref{fig:LS_overview}b).
Through microscopic one-way \textit{primary valves}, these \acp{VN} collect \textit{lymph}, a fluid comprised of plasma, macromolecules, and lymphocytes that leaks from capillaries, preventing excessive interstitial fluid accumulation~\cite{Breslin2014, Breslin2019}.
Initial lymphatics drain into \textit{collecting lymphatics}, larger vessels with both endothelial and smooth muscle layers, which are segmented into serially connected units called \textit{lymphangions}, separated by \textit{secondary valves}, see Fig.~\ref{fig:LS_overview}c). 
Phasic contractions of the smooth muscle cells triggered by action potentials drive lymph transport actively, enabling flow against adverse pressure gradients such as those encountered in the upright posture~\cite{Breslin2019}.
Pre-nodal collecting lymphatics deliver lymph to \textit{lymph nodes}, see Fig.~\ref{fig:LS_overview}d), where lymph contents encounter T, B, and natural killer immune cells and where the \ac{LS} and \ac{CVS} are directly interfaced via arterioles and \acp{HEV}~\cite{Cooper2015}, making lymph nodes critical immune monitoring sites at which antigens can be sampled, immune responses initiated, and immune cells exchanged between the two circulatory systems~\cite{Breslin2019}. 
A single post-nodal collecting lymphatic typically exits each node. 
Globally, the collecting lymphatics form a tree structure in which vessel diameters increase toward the root.
The largest vessels, termed \textit{trunks} (Fig.~\ref{fig:LS_overview}e)), drain into the left and right subclavian veins, closing the loop with the \ac{CVS}~\cite{Breslin2019}, see Fig.~\ref{fig:LS_overview}f).

Beyond the vascular side of the \ac{LS} lie the \textit{lymphoid organs}, which form the tissue-based functional hubs of the immune system. 
The \textit{primary lymphoid organs}, bone marrow and thymus, are where immune cells are produced and mature, while the \textit{secondary lymphoid organs}, lymph nodes, spleen, tonsils, and mucosa-associated lymphoid tissue, are where mature immune cells encounter antigens and mount responses~\cite{Breslin2019}.
Among these, lymph nodes are of particular relevance from an \ac{MC} perspective, as they are directly embedded in the collecting lymphatic \ac{VN} and thus naturally reachable by solutes transported through lymphangion chains.

\section{Envisioned Lymphatic MC Applications}\label{sec:Applications}
In the following, we propose potential diagnostic and therapeutic \ac{MC} applications in the \ac{LS} and highlight the need for mathematical models to enable their design.

\subsection{Targeted Drug Delivery to Lymph Nodes}
\begin{figure}
    \centering
    \includegraphics[width=\linewidth]{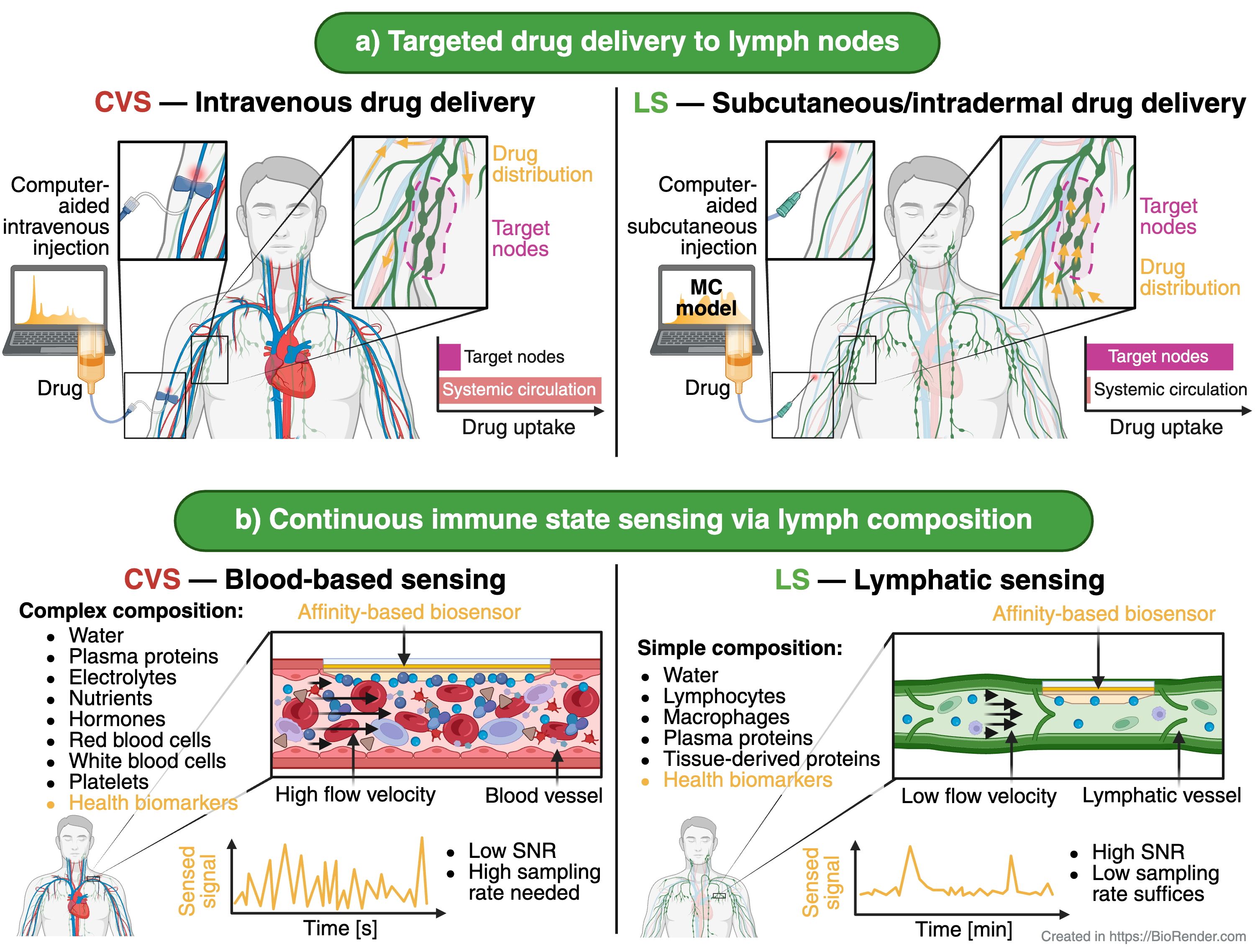}
    
    \caption{\textbf{Envisioned a) therapeutic and b) diagnostic applications of \ac{MC} in the \ac{LS}.} Comparisons to applications in the \ac{CVS} are shown.}\iffalse\caption{\textbf{Envisioned therapeutic and diagnostic applications of \ac{MC} in the \ac{LS}.} \textbf{a)}~Subcutaneous drug delivery exploits the natural lymphatic drainage anatomy to achieve more efficient drug uptake at target lymph nodes than intravenous delivery~\cite{Thomas2015}.
    \textbf{b)}~Immune state sensing via lymph composition potentially provides a higher-\ac{SNR} readout than blood-based sensing at a lower sampling rate, owing to the tissue-specific molecular signature of lymph and its biochemical simplicity relative to blood~\cite{Hansen2015}, and to comparatively low flow velocities.}\fi
    \label{fig:todo4}
\end{figure}

Lymph nodes are primary sites of immune activation and critical therapeutic targets in the treatment of cancer and \ac{HIV}, as well as in vaccine development; yet conventional intravenous drug delivery reaches them inefficiently since only a small fraction of a systemically circulating drug crosses from the bloodstream into the lymph node~\cite{Thomas2015}, cf.~Fig.~\ref{fig:todo4}a).
A more direct route exploits the natural drainage anatomy of the \ac{LS}: a drug injected subcutaneously or intradermally at a peripheral site, such as the hand, foot, or upper arm, is taken up by the initial lymphatics at that site and subsequently transported through a chain of collecting lymphatics and lymphangions toward the regional draining lymph node, e.g., the axillary or inguinal node~\cite{Breslin2019}, cf.~Fig.~\ref{fig:todo4}a).
However, the pulsatile, valve-gated nature of lymphangion flow strongly shapes the arrival timing and spatial distribution of molecules at the target node, see Section~\ref{sec:NumericalResults}, making dosage and release timing difficult to optimize without a quantitative transport model. 
An \ac{MC} framework, informed by such a model, could guide the design of injection protocols, including molecule size, dosage, and timing, that maximize local drug concentration at the target node while minimizing systemic exposure.
On the device side, this application requires either a simple injectable drug formulation with tuned physicochemical properties~\cite{Thomas2015}, or an implantable drug-release device capable of timed molecular release positioned along the afferent collecting lymphatic upstream of the target node.

\subsection{Continuous Immune State Sensing via Lymph Composition}
\iffalse
The molecular composition of lymph, i.e., the concentrations of cytokines, antigens, and metabolites, directly reflects the immune and inflammatory state of the upstream tissue it drains, making it an inherently informative signal for continuous health monitoring~\cite{Breslin2019}. 
An implantable \ac{MC} receiver positioned along a collecting lymphatic or within a lymph node could decode this signal to infer the tissue health state in near-real-time. 
Compared to blood-based sensing, the lymphatic environment offers two practical advantages: lymph flow is considerably slower~\cite{Bertram2013, Sharma2007}, giving molecules more time to interact with functionalized receptor surfaces, and lymph is biochemically simpler than blood, which carries a dense mixture of cells and proteins that raise the detection noise floor. 
This application requires a miniaturized, biocompatible implant with a functionalized sensing surface capable of selective molecular detection, analogous in principle to the biosensor technologies already employed in \textit{ex vivo} lymph node biopsy analysis, but adapted for continuous in-situ operation.
\fi

The molecular composition of lymph directly reflects the metabolic and immunological state of the upstream tissue it drains~\cite{Hansen2015, Breslin2019}: 
unlike blood plasma, which represents a systemic average, lymph carries a tissue-specific molecular signature enriched with cytokines, chemokines, extracellular matrix remodeling products, and markers of cellular apoptosis and necrosis~\cite{Hansen2015}. 
Proteomics studies have shown that lymph composition changes measurably and specifically in response to sepsis, acute inflammation, trauma, and infection, with biomarkers such as \ac{TNFalpha}, \ac{IL1beta}, and \ac{IL6} appearing at elevated concentrations in lymph~\cite{Hansen2015}, making it an inherently informative signal for continuous health monitoring. 
An implantable \ac{MC} receiver positioned along a collecting lymphatic or within a lymph node could decode this signal to infer the tissue health state in near-real-time, cf.~Fig.~\ref{fig:todo4}b). Compared to blood-based sensing, the lymphatic environment offers two practical advantages, cf.~Fig.~\ref{fig:todo4}b): 
First, lymph flow is considerably slower~\cite{Bertram2013, Sharma2007}, giving molecules more time to interact with functionalized receptor surfaces, thereby relaxing sampling-rate requirements for sensors.
Second, lymph is biochemically simpler than blood.
Many tissue-derived proteins and peptides are present in lymph at higher relative abundances than in blood plasma, or are lymph-specific altogether, as they are largely phagocytosed by local antigen-presenting cells before reaching venous circulation~\cite{Hansen2015}, resulting in a higher-\ac{SNR} sensing environment. 
Sensing applications in the \ac{LS} require a miniaturized, biocompatible implant with a functionalized sensing surface capable of selective molecular detection, cf.~Fig.~\ref{fig:todo4}b), analogous to biosensor technologies already employed in \textit{ex vivo} lymph node biopsy analysis, but adapted for continuous in-situ operation.
Crucially, quantitative transport models for lymphangion chains are needed to relate the signal detected at the sensor to the biomarker concentration at its tissue source, accounting for the transport delay and dispersion introduced by pulsatile, valve-gated flow.

\iffalse
\subsection{Early Detection of Infection and Inflammation}
When a pathogen enters peripheral tissue, the draining lymph node mounts an immune response within hours, and the molecular signature of this response (elevated cytokines and antigen fragments) propagates upstream through the afferent lymphatics.
An \ac{MC} sensor placed in a prenodal collecting vessel could detect this signal early, before systemic blood markers become elevated.
\fi

\subsection{Integration with Cardiovascular System Models}
The \ac{LS} and \ac{CVS} are tightly coupled: plasma continuously leaks from blood capillaries into the interstitium and is returned to the venous circulation via the thoracic and right lymphatic ducts, while \acp{HEV} in lymph nodes provide an exchange point between the two systems~\cite{Cooper2015}.
Holistic models of molecular signaling in the body, as relevant, e.g., to digital twin applications, cannot treat the two systems independently. 
A quantitative solute transport simulation of the \ac{LS} represents an initial step toward an interface between interstitial molecular dynamics and bloodborne \ac{MC} models, with the ultimate goal of enabling end-to-end quantification of how molecules secreted in peripheral tissue eventually reach the central circulation.

\section{Lymph Flow Model}\label{sec:LymphFlowModel}
Below, we summarize the \ac{0-D} lymph flow model of \cite{Bertram2010} and its numerical solution for computing flow rates, vessel diameters, and pressures. 
Rat and swine model parameters are listed in Table~\ref{tab:default_model_parameters}. 
For a detailed model description, cf.~\cite{Bertram2010, Bertram2013}.

\subsection{Model Equations}\label{ssec:Flow_Model_Equations}

\begin{figure}
    \centering
    \includegraphics[width=\linewidth]{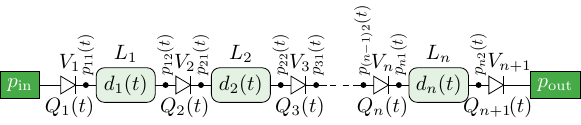}
    \caption{\textbf{Schematic of a chain of $n$ lymphangions~\cite{Bertram2010}.} Notation for valves, lymphangions, and lymph flow-related quantities is shown.}
    \label{fig:lymphangion_schematic}
\end{figure}

A chain of $n$ lymphangions is modeled as shown in Fig.~\ref{fig:lymphangion_schematic}.
Pressures $p_\mathrm{in}$ and $p_\mathrm{out}$ are prescribed at the chain inlet and outlet, respectively, with $p_\mathrm{out}>p_\mathrm{in}$, giving rise to the adverse pressure gradient $\Delta P=p_\mathrm{out}-p_\mathrm{in}>0$.
Lymphangion $L_i$, $i\in\{1,\ldots,n\}$, has common length $l$, diameter $d_i(t)$, rest diameter $\bar{d}_i$, and wall stiffness $P_{i}$.
The pressures at its up- and downstream ends are denoted by $p_{i1}(t)$ and $p_{i2}(t)$, respectively.
$n+1$ pressure-gated valves $V_j$, $j\in\{1,\ldots ,n+1\}$, are located between adjacent lymphangions and at the inlet and outlet of the chain, cf.~Fig.~\ref{fig:lymphangion_schematic}.
Depending on the valve resistance $R_{j}$, which is dictated by the transvalvular pressure difference $\Delta p_j(t)=p_{(j-1)2}(t)-p_{j1}(t)$, a positive or negative flow rate $Q_j(t)$ passes through valve $V_j$.
Pulsatile lymph flow is driven by periodic smooth muscle cell contractions with frequency $f$, strength $M$, lymphangion-specific time offsets $t_{i0}$, and refractory period $t_r$.
An external pressure $p_e$ acts on all lymphangions, resulting in the transmural pressure $p_{\mathrm{tm},i}$ of lymphangion $L_i$.

\renewcommand{\arraystretch}{1.35}
\begin{table}
    \centering
    \caption{\textbf{Default parameter values for rat~\cite{Bertram2010, Bertram2013} and swine~\cite{Sharma2007}} given in the \ac{CGS} system of units.}\vspace*{1mm}
    \resizebox{\linewidth}{!}{%
    \begin{tabular}{|c||l|c|c|c|}
        \hline
        \rowcolor{mydarkgreen}
        \textcolor{white}{\textbf{Parameter}} & \textcolor{white}{\textbf{Description}} & \textcolor{white}{\textbf{Rat~\cite[Table~I, V3]{Bertram2010}}} & \textcolor{white}{\textbf{Swine (based on~\cite{Sharma2007})}} & \textcolor{white}{\textbf{Unit}} \\
        \hline\hline
        $n$ & Number of lymphangions & 4 & 12 & -- \\ \cline{1-5}
        $\bar{d}_{i}$ & Rest diameter of $L_i$ & $\{0.025, 0.022, \ldots, 0.016\}$ & $\{0.2381, 0.2303, \ldots, 0.1524\}$ & $\SI{}{\centi\meter}$ \\ \cline{1-5}
        $l$ & Lymphangion length & $0.3$ & $1.04$ & $\SI{}{\centi\meter}$ \\ \cline{1-5}
        $p_\mathrm{in}$, $p_\mathrm{out}$ & Inlet and outlet pressure & $2275$, $2875$ & $2275$, $2475$ & $\SI{}{\dyne\per\centi\meter\squared}$ \\ \cline{1-5}
        $\Delta P$ & Adverse pressure & $600$ & $200$ & $\SI{}{\dyne\per\centi\meter\squared}$ \\ \cline{1-5}
        $p_\mathrm{e}$ & External pressure & $2100$ & $2100$ & $\SI{}{\dyne\per\centi\meter\squared}$ \\ \cline{1-5}
        $p_\mathrm{open}$ & Valve opening pressure & $-70$ & $-70$ & $\SI{}{\dyne\per\centi\meter\squared}$ \\ \cline{1-5}
        $s_\mathrm{open}$ & Valve opening slope & $0.04$ & $0.04$ & $\SI{}{\centi\meter\squared\per\dyne}$ \\ \cline{1-5}
        $R_{\mathrm{min}/\mathrm{max}}$ & Min./max.~valve resistance & $600$, $\SI{1.2e7}{}$ & $600$, $\SI{1.2e7}{}$ & $\SI{}{\dyne\second\per\centi\meter\squared\per\milli\liter}$ \\
        \cline{1-5}
        $\mu$ & Lymph viscosity & $1$ & $1$ & $\SI{}{\centi\poise}$ \\ \cline{1-5}
        $P_{i}$ & Passive wall stiffness ($L_i$) & $\{50, 75, \ldots, 125\}$ & $\{50, 56.818, \ldots, 125\}$ & $\SI{}{\dyne\per\centi\meter\squared}$ \\ \cline{1-5}
        $f$ & Contraction frequency & $0.5$ & $0.16$ & $\SI{}{\hertz}$ \\ \cline{1-5}
        $M$ & Contraction strength & $3.6$ & $34$ & $\SI{}{\dyne\per\centi\meter}$ \\ \cline{1-5}
        $t_{i0}$ & Contraction time offset ($L_i$) & $i/2$ & $i/0.64$ & $\SI{}{\second}$ \\ \cline{1-5}
        $t_r$ & Refractory period & $0$ & $36$ & $\SI{}{\second}$ \\ 
        \cline{1-5}
        $D$ & Molecular diffusion coeff. & $\SI{1e-6}{}$ & $\SI{1e-6}{}$ & $\SI{}{\centi\meter\squared\per\second}$ \\ 
        \cline{1-5}
        $l_{\mathrm{Rx}_v}$ & Length of $\mathrm{Rx}_v$ & $\SI{0.05}{}$ & $\SI{0.35}{}$ & $\SI{}{\centi\meter}$ \\ 
        \hline
    \end{tabular}
    }
    \label{tab:default_model_parameters}
    \vspace*{-4mm}
\end{table}
\renewcommand{\arraystretch}{1}

In lymphangion $L_i$, fluid mass is conserved according to~\cite{Bertram2010}
\begin{align}\label{eqn:conservation_of_fluid_mass}
    Q_{i+1}(t)=Q_i(t)-\frac{\pi}{2}d_i(t)l\frac{\mathrm{d}d_i(t)}{\mathrm{d}t},
\end{align}
and the conservation of fluid momentum under assumption of laminar flow is captured by~\cite{Bertram2010}
\begin{align}\label{eqn:conservation_of_fluid_momentum}
    \frac{p_{i1}(t)-p_{i2}(t)}{l}=\frac{64\mu(Q_i(t)+Q_{i+1}(t))}{\pi d_i(t)^4},
\end{align}
where $\mu$ denotes the lymph viscosity.
The transmural pressure $p_{\mathrm{tm},i}$ in lymphangion $L_i$ is defined as~\cite{Bertram2010}
\begin{align}\label{eqn:Constitutive_Relation}
    p_{\mathrm{tm},i}(t) &=\frac{p_{i1}(t)+p_{i2}(t)}{2} - p_\mathrm{e}\nonumber\\ 
    &= P_{i}\left[\exp \left(\frac{d_i(t)}{\bar{d}_{i}}\right)
    - \left(\frac{\bar{d}_{i}}{d_i(t)}\right)^3\right] + \frac{M_i(t)}{d_i(t)},
\end{align}
where the latter sum term models (phasic) muscle contractions along the
chain of lymphangions, where~\cite{Bertram2013}
\begin{align}\label{eqn:refractory}
    \hspace*{-2mm}M_i(t) \hspace*{-.5mm}=\hspace*{-.5mm}
    \begin{cases}
        \hspace*{-.5mm}M\left(1\hspace*{-.5mm}-\hspace*{-.5mm}\cos(2\pi f\,\tau_i(t))\right), & \hspace*{-2mm}0\hspace*{-.5mm} \le\hspace*{-.5mm} \tau_i(t)\hspace*{-.5mm} <\hspace*{-.5mm} 1/f, \\
        \hspace*{-.5mm}0, & \hspace*{-2mm}1/f \hspace*{-.5mm}\le\hspace*{-.5mm} \tau_i(t) \hspace*{-.5mm}< \hspace*{-.5mm}1/f \hspace*{-.5mm}+\hspace*{-.5mm} t_\mathrm{r},
    \end{cases}
\end{align}
and $\tau_i(t) = (t - t_{i0}) \bmod (1/f + t_\mathrm{r})$ is the time elapsed since
$L_i$'s most recent contraction began. Here, $\bmod$ denotes the modulo operator. During the refractory period $t_r$, no further contraction can be initiated;
the total pumping period is thus $1/f + t_\mathrm{r}$.
Lastly, the resistance of valve $V_j$ is given as~\cite{Bertram2010}
\begin{align}\label{eqn:valve_resistance}
    R_{j}(\Delta p_j,t)=R_{\mathrm{min}}\hspace*{-.5mm}+\hspace*{-.5mm}\frac{R_{\mathrm{max}}}{1\hspace*{-.5mm}+\hspace*{-.5mm}\exp(s_\mathrm{open}(\Delta p_j(t)\hspace*{-.5mm}-\hspace*{-.5mm}p_\mathrm{open}))},
\end{align}
where $R_{\mathrm{min}}$, $R_{\mathrm{max}}$, and $s_\mathrm{open}$ denote the minimum and maximum valve resistance, and the valve opening slope, respectively.
Note that, in contrast to~\cite{Bertram2010}, valve failure caused by very large adverse pressure gradients is neglected in~\eqref{eqn:valve_resistance}, as the pressure differences considered in this work remain well below the failure threshold reported in~\cite{Bertram2010}.

\subsection{Solving the Model Equations}

The system in Subsection~\ref{ssec:Flow_Model_Equations} couples a non-linear \ac{ODE} in $d_i(t)$, cf.~\eqref{eqn:conservation_of_fluid_mass}, to an algebraic relation for pressures and flow rates, cf.~\eqref{eqn:conservation_of_fluid_momentum}. 
The only initial conditions required are the diameters $d_i(0) = \bar{d}_{i}$, cf.~Table~\ref{tab:default_model_parameters}, corresponding to the passive rest state of each lymphangion; all pressures and flow rates at $t=0$ are then determined algebraically via~\eqref{eqn:conservation_of_fluid_momentum}.
Subsequently, for $t>0$, at each time step, given $d_i$ and $t$, the relation in~\eqref{eqn:Constitutive_Relation} directly yields the pressure sum $s_i \defeq p_{i1}+p_{i2} = 2(p_e + p_{\mathrm{tm},i})$ in each lymphangion. 
Taking $p_{i1},i\in\{1,\ldots,n\}$, as $n$ unknowns, the downstream pressures follow as $p_{i2} = s_i - p_{i1}$, and the transvalvular pressure differences $\Delta p_j$ are then fully determined. 
The $n+1$ valve flow rates $Q_j = \Delta p_j / R_{\mathrm{v},j}(\Delta p_j)$, expressed using the valve resistances in~\eqref{eqn:valve_resistance}, are substituted into the momentum equation~\eqref{eqn:conservation_of_fluid_momentum}, yielding $n$ nonlinear residual equations in $p_{i1}$ that are solved by Newton iteration, warm-started from the previous time step. 
The resulting flow rates enter the right-hand side of the mass conservation equation~\eqref{eqn:conservation_of_fluid_mass}, which is integrated forward in time by SciPy's \textit{Radau} solver.

\section{Particle-Based Simulation Framework for Solute Transport}\label{sec:MoleculeTransportSimulation}
\begin{figure}
    \centering
    \includegraphics[width=\linewidth]{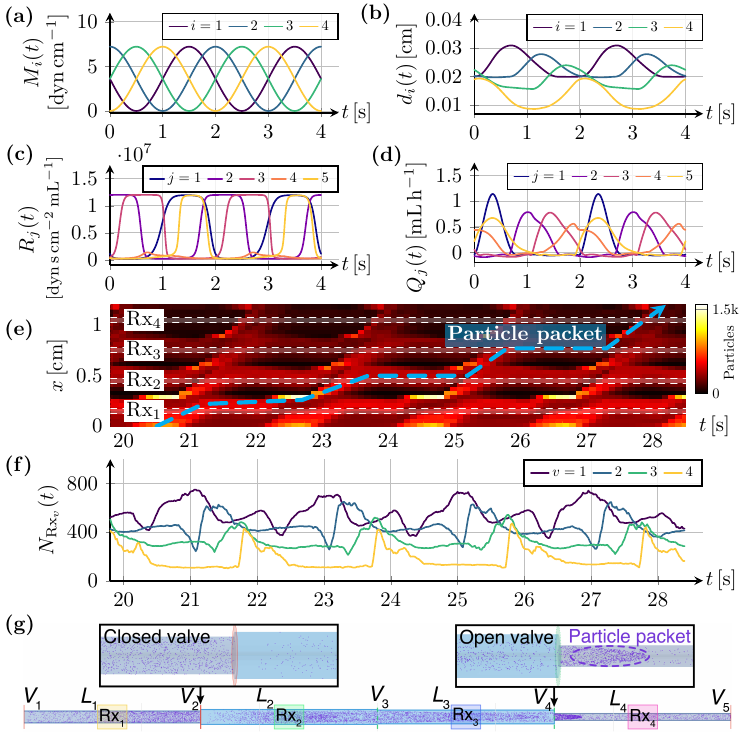}
    \caption{\textbf{\ac{PBS} of solute transport in a rat.} Default rat parameters apply, cf.~Table~\ref{tab:default_model_parameters}. \textbf{a)--d)} Lymph flow model~\cite{Bertram2010,Bertram2013}. \textbf{e)--g)} \ac{PBS}.}
    \label{fig:default_rat}
\end{figure}

Below, we propose a \ac{PBS} for simulating solute transport in multi-lymphangion lymphatic vessels, taking the predictions of the flow model in Section~\ref{sec:LymphFlowModel} as input. 
Particle propagation is modeled in \ac{3-D} space, cf.~Fig.~\ref{fig:default_rat}g), with $x$ denoting the longitudinal dimension of the chain and $y$, $z$ spanning the vessel cross-section.
Given the low Reynolds numbers of lymph flow~\cite{Bertram2010,Breslin2019}, we assume a time-varying, laminar velocity profile~\cite{Medina2022} within any lymphangion $L_i$, driven by the spatially averaged flow rate $\bar{Q}_i(t)=\frac{1}{2}(Q_i(t)+Q_{i+1}(t))$, yielding
\begin{equation}
    u(r,t)=
        \begin{cases}
        \frac{2\bar{Q}_i(t)}{A(t)}\left(1-\frac{r^2}{\frac{1}{4}d_i^2(t)}\right), &\text{ $r<\frac{1}{2}d_i(t)$},\\
        0,&\text{}\text{ $r\geq\frac{1}{2}d_i(t)$},
        \end{cases}
\end{equation}
where $A(t)=\frac{1}{4}\pi d_i^2(t)$ is the cross-sectional area and $r=\sqrt{y^2+z^2}$ the radial distance from the cross-sectional center.

Particles are assumed sufficiently dilute to leave the fluid mechanics unaffected and are therefore modeled as passive tracers. 
Since the initial lymphatics are not explicitly modeled here, we approximate them as a well-mixed upstream reservoir with fixed particle concentration $c_0$ before the first lymphangion $L_1$, from which tracers are drawn into the chain at the rate $c_0Q_1(t)$ starting at $t=0$ with default concentration $c_0=\SI{2e7}{\particle\per\milli\liter}$, chosen to ensure limited counting noise. 
Because this rate is tied to the flow rate, release only occurs while valve $V_1$ admits forward flow ($Q_1(t)>0$).

Within each time step of size $\Delta t=\SI{5e-3}{\second}$, all time-varying quantities are updated as follows:
First, $d_i(t)$ and $Q_i(t)$ are sampled at the start of the step.
Second, particles are released from the reservoir into $L_1$.
Third, particle positions are updated due to advective--diffusive transport according to 
\begin{align}
    &x\leftarrow x + u(r,t)\Delta t+\varepsilon_x,
    &&y\leftarrow y+\varepsilon_y, &z\leftarrow z+\varepsilon_z,
\end{align}
with \ac{IID} \acp{RV} $\varepsilon_x,\varepsilon_y,\varepsilon_z\sim\mathcal{N}(0,2D\Delta t)$, redrawn at each time step, where $\mathcal{N}(a,b)$ denotes the normal distribution with mean $a$ and variance $b$.
Fourth, particles that have left the vessel domain due to diffusion or shrinking vessel diameters, are reflected about the current wall radius, preserving the azimuthal angle.
Fifth, particles are reflected at valves if the instantaneous flow through the valve does not point in the direction they are attempting to cross, i.e., $Q_i(t)\le0$ for a particle attempting to move downstream through valve $V_i$, or $Q_i(t)\ge0$ for one attempting to move upstream\footnote{Although the valve resistance varies continuously with the transvalvular pressure difference $\Delta p_i(t)$ in the flow model, cf.~\eqref{eqn:valve_resistance}, valves are treated as binary for particles: particles either pass with the flow or are reflected. This neglects finer-scale flow structures that may occur in partially open valves.}.
Sixth, particles that have left the vessel after crossing via a valve into a lymphangion, whose current radius is smaller than the particle's radial position, are reflected back into the vessel.
Seventh, particles that have exited the chain (past the outlet at $x=nl$, or past the inlet during backflow) are flagged as \textit{exited} and excluded from further updates. 
Lastly, the simulation time is advanced, i.e., $t\leftarrow t+\Delta t$.
To track particle numbers over time, we employ transparent \acp{Rx}. Each $\mathrm{Rx}_v$, $v\in\mathbb{N}$, has length $l_{\mathrm{Rx},v}$, is centered at $x_{\mathrm{Rx},v}$, spans the full cross-section, and counts the number of particles $N_{\mathrm{Rx}_v}(t)$ in its volume.

\section{Simulation Results}\label{sec:NumericalResults}

In the following, we investigate the proposed \ac{PBS} framework by analyzing the impact of selected system parameters on solute transport using the rat parameters in Table~\ref{tab:default_model_parameters} and subsequently validate the model against \textit{in vivo} data from the swine model in~\cite{Sharma2007} for the swine parameters in Table~\ref{tab:default_model_parameters}.

\subsection{Parameter Effects on Pumping and Particle Transport}

Fig.~\ref{fig:default_rat} summarizes pumping behavior and particle transport in a rat lymphangion chain of length $n=4$~\cite{Bertram2010}, with Fig.~\ref{fig:default_rat}g) providing a snapshot of the \ac{PBS} (cf. also~\cite[\texttt{video1}]{jakumeit2026zenodo}).
Pumping is driven by phasic muscle contractions (active tension $M_i(t)$, Fig.~\ref{fig:default_rat}a)), which, together with the passive wall stiffness $P_i$ and rest diameters $\bar{d}_i$ of the lymphangions $L_i$, produce the time-varying diameters $d_i(t)$ in Fig.~\ref{fig:default_rat}b).
These diameters vary inversely with the contractions and decrease along the chain, following the decreasing $\bar{d}_i$ (Table~\ref{tab:default_model_parameters}).
The valve resistances $R_i(t)$ in Fig.~\ref{fig:default_rat}c) show that $V_1$, $V_2$, $V_3$, and $V_5$ each swing fully shut and open once per contraction, whereas $V_4$ never faces enough adverse pressure to reach its maximum resistance.
Correspondingly, the flow rates $Q_i(t)$ in Fig.~\ref{fig:default_rat}d) are predominantly positive, with only small backflows between forward pulsations, indicating effective pumping for the rat parameters in Table~\ref{tab:default_model_parameters}.
Note that Figs.~\ref{fig:default_rat}a)–c) reproduce Fig.~5 in~\cite{Bertram2010}, confirming the correct implementation of the flow model.

Fig.~\ref{fig:default_rat}e) illustrates the resulting spatiotemporal particle transport, which proceeds in clear, step-like movements of particle packets, i.e., spatially localized clusters of particles, through the chain (cf.~cyan arrow): particles are repeatedly pushed against a closed valve, temporarily compressing into a dense packet (left zoom-in, Fig.~\ref{fig:default_rat}g) and \cite[\texttt{video4}]{jakumeit2026zenodo}), before being ejected once the valve reopens (right zoom-in, Fig.~\ref{fig:default_rat}g) and \cite[\texttt{video4}]{jakumeit2026zenodo}). 
This compression-and-release cycle is consistent with other models, cf., e.g., \cite[Fig.~2]{Li2024}, and is the key mechanism underlying the \textit{bursty} received signals in Fig.~\ref{fig:default_rat}f), with peak-to-peak times set by the pumping period $1/f + t_\mathrm{r} = \SI{2}{\second}$. 
Signals become progressively sharper further along the chain (e.g., $N_{\mathrm{Rx}_1}(t)$ vs.\ $N_{\mathrm{Rx}_4}(t)$), as repeated compression at successive valves sharpens the particle packets.

\begin{figure}
    \centering
    \includegraphics[width=\linewidth]{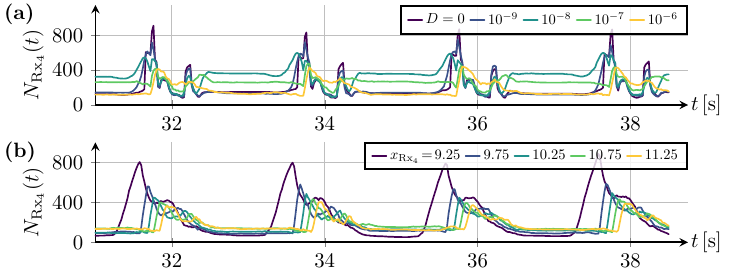}
    \caption{\textbf{Impact of molecular diffusion coefficient $D$ and \ac{Rx} position $x_\mathrm{Rx}$ within a lymphangion.} Default rat parameters apply, cf.~Table~\ref{tab:default_model_parameters}. $D$ and $x_\mathrm{Rx}$ are given in $\SI{}{\centi\meter\squared\per\second}$ and $\SI{}{\milli\meter}$, respectively.}
    \label{fig:parameter_sweep_1}
\end{figure}

Fig.~\ref{fig:parameter_sweep_1} further examines how the molecular diffusion coefficient $D$ and \ac{Rx} position within a lymphangion shape the received signal (cf.~also \cite[\texttt{video2}]{jakumeit2026zenodo}).
Decreasing $D$ leads to several distinct peaks per pumping period, which smear into a single broader peak as $D$ increases. 
Similarly, a \ac{Rx} placed immediately after a valve (e.g., $x_\mathrm{Rx}=9.25$) sees substantially larger peak amplitudes than one placed farther away (e.g., $x_\mathrm{Rx}=11.25$), since particle packets are densest immediately upon discharge from a valve and disperse via diffusion as they travel along the lymphangion, cf. also~\cite[\texttt{video4}]{jakumeit2026zenodo}.

\begin{figure}
    \centering
    \includegraphics[width=\linewidth]{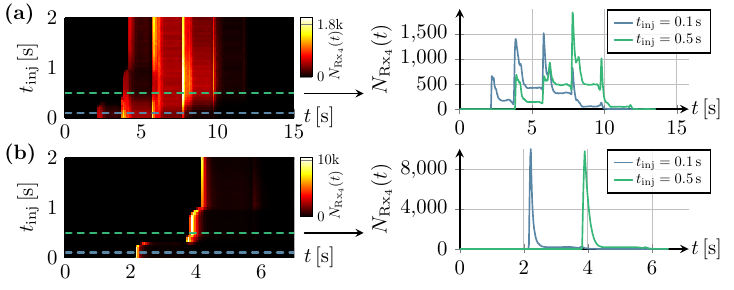}
    \caption{\textbf{Impact of bolus release type and injection time $t_\mathrm{inj}$.} Default rat parameters apply, cf.~Table~\ref{tab:default_model_parameters}. \textbf{a)}~Uniform bolus injection in cross-section. \textbf{b)}~Point release in cross-sectional center. For both release types, $20000$ particles are released at the longitudinal center of the first lymphangion $L_1$ ($x=\SI{0.15}{\centi\meter}$). The signal $N_{\mathrm{Rx}_4}(t)$ is color-coded and zoom-ins for $t_\mathrm{inj}=\{\SI{0.1}{\second},\SI{0.5}{\second}\}$ are shown.}
    \label{fig:parameter_sweep_2}
    \vspace*{2mm}
\end{figure}

While a continuous release is assumed throughout the paper (cf.~Section~\ref{sec:MoleculeTransportSimulation}), Fig.~\ref{fig:parameter_sweep_2} instead probes the time-varying channel with an instantaneous bolus injection, plotting the received particle count $N_{\mathrm{Rx}_4}(t)$ against both $t$ and the injection time $t_\mathrm{inj}$ within one pumping period (cf. also~\cite[\texttt{video5}]{jakumeit2026zenodo}). 
Particles are released uniformly across the cross-section in Fig.~\ref{fig:parameter_sweep_2}a) and as a point release at the cross-sectional center in Fig.~\ref{fig:parameter_sweep_2}b). 
In the uniform case, a single bolus produces multiple peaks at $\mathrm{Rx}_4$, as successive valve closures repeatedly split the particle packet; notably, these peaks occur at nearly the same $t$ regardless of $t_\mathrm{inj}$, i.e., the valves "synchronize" particles injected at different times.
The point release instead produces only one major peak for any $t_\mathrm{inj}$: released on the channel centerline, particles experience much higher flow velocities, so the packet stays compact en route to $\mathrm{Rx}_4$ and is never split up by a valve.

\subsection{Simulation Validation Against In-Vivo Imaging}

\begin{figure}
    \centering
    \includegraphics[width=\linewidth]{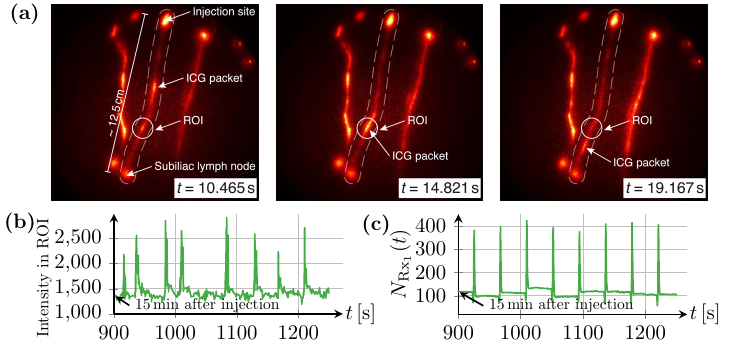}
    \caption{\textbf{\ac{PBS} validation with \textit{in vivo} fluorescence imaging data from a swine model.} \textbf{a)} Snapshots taken from~\cite{Sharma2007}. \textbf{b)} \textit{In vivo} data were recovered from Fig.~4 in~\cite{Sharma2007} using WebPlotDigitizer (\url{https://automeris.io/}). \textbf{c)} \ac{PBS} data uses the swine parameters in Table~\ref{tab:default_model_parameters}.}
    \label{fig:validation}
\end{figure}

Lastly, we compare the predictions of the \ac{PBS} with \textit{in vivo} measurements of fluorescent tracer transport in collecting lymphatic vessels, obtained from the abdomen of a Yorkshire swine model\footnote{The swine abdomen closely resembles the human lymphatic plexus~\cite{Sharma2007}.} and reported in~\cite{Sharma2007}. 
In these experiments, $\SI{200}{\micro\liter}$ of $\SI{32}{\micro\molar}$ \ac{ICG} was injected via a catheter at each of four sites, draining into distinct multi-lymphangion collecting vessels of approximately $\SI{12.5}{\centi\meter}$ in length. 
The recorded videos show a steadily sustained supply of \ac{ICG} entering the vessel, consistent with tracer being gradually collected from the subcutaneous depot into the initial lymphatics and mirroring the continuous particle release used in the \ac{PBS} in Section~\ref{sec:MoleculeTransportSimulation}. 
This enables a direct comparison between simulated and observed transport.
We consider the second vessel from the left (Fig.~\ref{fig:validation}a), which drains into the subiliac lymph node located $2.5$--$\SI{3}{\centi\meter}$ below the epidermis~\cite{Sharma2007}. 
In\cite{Sharma2007}, tracer propagation was recorded using near-infrared fluorescence imaging together with hematoxylin and eosin micrographs, allowing individual \ac{ICG} packets to be tracked as they transit the vessel over several minutes; representative still frames are shown in Fig.~\ref{fig:validation}a). To quantify the spatiotemporal tracer distribution, the pixel intensity over time within a \ac{ROI} placed on the vessel was measured~\cite{Sharma2007}, as shown in Fig.~\ref{fig:validation}b).
In the \ac{PBS}, $\mathrm{Rx}_1$ is placed at the same relative position along the lymphangion chain as in the measurements in~\cite{Sharma2007} and $N_{\mathrm{Rx}_1}(t)$ is evaluated for reservoir concentration\footnote{Chosen for simulation reasons, not physically motivated.} $c_0=\SI{2e4}{\particle\per\milli\liter}$, cf.~Fig.~\ref{fig:validation}c) and \cite[\texttt{video6}]{jakumeit2026zenodo}.
The swine parameters in Table~\ref{tab:default_model_parameters} were chosen to match available data in~\cite{Sharma2007} where possible, retained from \cite{Bertram2010} where species-invariant, and otherwise scaled anatomically. 
Specifically, $n$ and $l$ reproduce the $\SI{12.5}{\centi\meter}$ vessel length in \cite{Sharma2007}, while $f$ and $t_\mathrm{r}$ reproduce its reported \ac{ICG} pulse frequency of roughly $1.5$ pulses/min. 
Valve gating parameters and $\mu$ were retained from \cite{Bertram2010,Bertram2013} as fluid and valve microstructure properties are expected to be conserved across species; remaining structural parameters were scaled to reflect the larger anatomical scale of swine lymphatics.

Comparing Figs.~\ref{fig:validation}b) and \ref{fig:validation}c), we observe that the \ac{PBS} reproduces the bursty characteristic of the \textit{in vivo} signal, with a similar order of inter-peak times ($\approx\SI{42}{\second}$). 
However, inter-peak times vary more strongly \textit{in vivo} (Fig.~\ref{fig:validation}b)) than in the \ac{PBS} (Fig.~\ref{fig:validation}c)) where they are essentially deterministic, i.e., set by the pumping period $1/f + t_\mathrm{r}$, aside from minor fluctuations due to diffusion. 
This additional variability \textit{in vivo} likely stems from irregular muscle contractions along the lymphangions and from potentially varying rates of \ac{ICG} transfer from the interstitial tissue into the lymphatic vessel. 
Peak amplitudes show a similar pattern: they fluctuate more strongly \textit{in vivo}, whereas in the \ac{PBS} their variation is likewise attributable to diffusion.
Overall, the \ac{PBS} captures the key qualitative signal characteristics observed \textit{in vivo}, with the remaining discrepancies pointing to physiological variability not represented in the model.

\section{Conclusion}\label{sec:Conclusion}
In this paper, we presented the first focused treatment of the \ac{LS} from an \ac{MC} perspective, proposing envisioned health-related \ac{MC} applications alongside a \ac{PBS} framework for particle transport due to the pulsatile flow of a multi-lymphangion lymphatic vessel. 
We argued that, compared to the \ac{CVS}, where most therapeutic and diagnostic \ac{MC} applications are envisioned, the \ac{LS} is preferable for targeted drug delivery to lymph nodes and for continuous immune state sensing. 
The pulsatile lymph flow in the \ac{PBS} was informed by the established model in~\cite{Bertram2010, Bertram2013} and validated against \textit{in vivo} measurements of \ac{ICG} propagation in an abdominal lymphatic vessel of a Yorkshire swine model~\cite{Sharma2007}, showing good agreement in qualitative signal characteristics. 
We then used the \ac{PBS} to investigate how key system parameters influence lymphangion pumping dynamics and the resulting particle transport, demonstrating that the valve-gated nature of lymph flow causes bursty received signals.

Future work could extend the \ac{PBS} to include initial lymphatics, lymph nodes, and lymphatic \acp{VN}, enabling more realistic simulations of solute transport in the \ac{LS}. 
The \ac{PBS} and the experimental measurements from~\cite{Sharma2007} could further serve as the basis for deriving an analytical transport model, which could aid in the design of the applications proposed in Section~\ref{sec:Applications}.

\bibliography{bibliography}

\end{document}